\documentclass[prd, lettersize,twocolumn,superscriptaddress,
preprintnumbers,nofootinbib,showpacs]{revtex4-1}
\usepackage{amsmath}
\usepackage{physics}
\usepackage{graphicx}
\usepackage{amssymb}
\usepackage[colorlinks,citecolor=red]{hyperref}

\usepackage{color}    
\usepackage{ulem}
\usepackage{multirow}
\usepackage{slashed}

\newcommand{\beq}{\begin{equation}}
\newcommand{\eeq}{\end{equation}}
\newcommand{\bea}{\begin{eqnarray}}
\newcommand{\eea}{\end{eqnarray}}
\newcommand{\nn}{\nonumber}

\begin{document}

\title{Probing the  $H\gamma\gamma$ coupling via Higgs exclusive decay into quarkonia plus a photon at the HL-LHC}

\author{Hongxin Dong}
\email{211002010@njnu.edu.cn}
\affiliation{Department of Physics and Institute of Theoretical Physics, Nanjing Normal University, Nanjing, Jiangsu 210023, China}
\affiliation{Institute of Modern Physics, Chinese Academy of Sciences, Lanzhou, Gansu Province 730000, China}

\author{Peng Sun}
\email{pengsun@impcas.ac.cn}
\affiliation{Institute of Modern Physics, Chinese Academy of Sciences, Lanzhou, Gansu Province 730000, China}

\author{Bin Yan}
\email{yanbin@ihep.ac.cn (corresponding author)}
\affiliation{Institute of High Energy Physics, Chinese Academy of Sciences, Beijing 100049, China}

\date{\today}

\begin{abstract}
It is well known that the precise measurement of $H\to\gamma\gamma$ can generate two degenerate parameter spaces of $H\gamma\gamma$ anomalous coupling.
We propose to utilize the exclusive Higgs rare decays $H\to\Upsilon(ns)+\gamma$ to break above mentioned degeneracy and further to constrain the $H\gamma\gamma$ anomalous coupling at the HL-LHC. We demonstrate that the branching ratios of $H\to\Upsilon(ns)+\gamma$ can be significantly enhanced in the non-SM-like parameter space from $H\to\gamma\gamma$ measurement, due to the destructive interference between the direct and indirect production of $\Upsilon(ns)$ in the SM. By applying the NRQCD factorization formalism, we calculate the partial decay widths of $H\to\Upsilon(ns)+\gamma$ at the NLO accuracy of $\alpha_s$. We show that it is hopeful to break such degenerate parameter space at the HL-LHC if we can further highly suppress the background and enhance the signal efficiency compared to the ATLAS preliminary simulation.
\end{abstract}

\maketitle

\section{Introduction}
After the discovery of the Higgs boson at the Large Hadron Collider (LHC)~\cite{ATLAS:2012yve,CMS:2012qbp}, precise measuring the Higgs properties at the LHC and future colliders has became
one of the major tasks of particle physics. The rare decay process, $H\to \gamma\gamma$, as one of the golden channel to discover the Higgs boson
has received much attention in the high energy physics community due to the excellent performance of photon reconstruction and identification at the LHC. 
 It is useful to indirectly search the potential new physics (NP) effects through measuring the $H\gamma\gamma$ anomalous coupling
 if the NP contributes to this rare decay by quantum loops. From the recent global analysis of the ATLAS collaboration, the signal strength of $\gamma\gamma$ mode is consistent with the SM prediction with a high accuracy, therefore, the $H\gamma\gamma$ anomalous coupling has been severely constrained by the LHC data~\cite{ATLAS:2022tnm}.
 
However, all the knowledge of the $H\gamma\gamma$ anomalous coupling is inferred from the Higgs decay branching ratio measurements, as a result, it would be a challenge to probe the NP effects in the so called ``faked-no-new-physics" (FNNP) scenario that the NP contribution is about minus two times the SM contribution~\cite{Cao:2015fra,Cao:2015iua,Cao:2015oaa,Cao:2016zob,Li:2019uyy,Yan:2021veo,Yan:2021htf,Li:2021uww,Dong:2022ayy}.
In such case, the branching ratio of $H\to\gamma\gamma$ would be similar to the Standard Model (SM) prediction and this scenario can not be distinguished from the SM through the Higgs decays to $\gamma\gamma$ mode at colliders.  One of the approach to break the above mentioned degeneracy is trying to measure the cross section of $e^+e^-\to H\gamma$ at lepton colliders~\cite{Cao:2015fra,Cao:2015iua,Li:2015kxc,Sang:2017vph,Demirci:2019ush}. It arises from the fact that the interference effects between the diagrams which involve the $H\gamma\gamma$ anomalous coupling and the other diagrams, and  the energy dependence of the cross section will shift the dependence of the $H\gamma\gamma$ anomalous coupling~\cite{Cao:2015fra,Cao:2015iua}.
But, both the $H\gamma\gamma$ and $HZ\gamma$ anomalous couplings could contribute to the cross section of $e^+e^-\to H\gamma$, therefore,  we can not pin down the magnitude and sign of  $H\gamma\gamma$ and $HZ\gamma$ anomalous couplings separately from this measurement.

\begin{figure}
\centering
\includegraphics[scale=0.32]{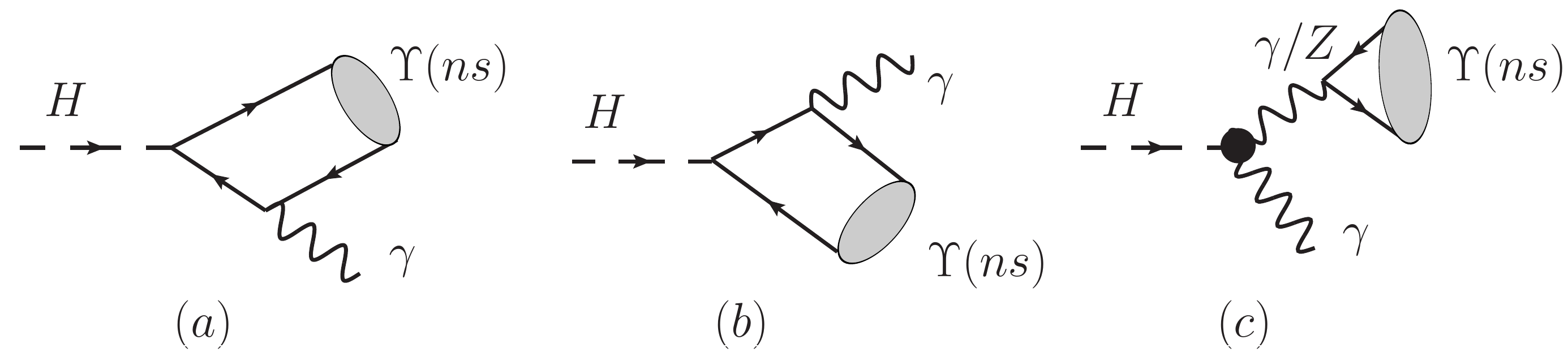}
\caption{The leading order Feynman diagrams of $H\to \Upsilon(ns)+\gamma$. The black dot denotes the effective $H\gamma\gamma$ and $HZ\gamma$ couplings from the SM and NP. }
\label{Fig:FeyLO}
\end{figure}

In this work, we propose yet another novel idea to probe the FNNP scenario  at the high luminosity LHC (HL-LHC) through the exclusive Higgs boson decay process $H\to \Upsilon(ns)+\gamma$, with $n=1,2,3$. There are two separate production mechanisms for the quarkonium state: the direct production (see Fig.~\ref{Fig:FeyLO}(a),(b)) and the indirect production (see Fig.~\ref{Fig:FeyLO}(c)).
The indirect production of the $\Upsilon(ns)$ in this process can be induced by the top quark and $W$-boson loop in the SM.
It has been shown that  this process can be used to constrain the light quark Yukawa coupling and its CP violation effects via the interference between the direct and indirect process~\cite{Bodwin:2013gca,Bodwin:2014bpa,Kagan:2014ila,Konig:2015qat,Modak:2016cdm,Zhou:2016sot,Bodwin:2016edd,Brambilla:2019fmu}.  In this paper, we demonstrate that this exclusive Higgs decay can also be used to probe the FNNP scenario at the HL-LHC. 

Owing to the destructive interference between the direct and indirect production of the process $H\to \Upsilon(ns)+\gamma$ in the SM, the branching ratio of this rare decay is around $\mathcal{O}(10^{-9})\sim \mathcal{O}(10^{-8})$~\cite{Bodwin:2013gca,Bodwin:2014bpa,Konig:2015qat}. Thus, it would be a challenge to observe the SM signal with the current or foreseeable dataset. But, the branching ratio could be enhanced by one  
to two orders of magnitude for the FNNP scenario due to the change of the sign of the interference term between the direct and indirect production processes
\footnote {The branching ratio of the charmonia final state is determined mainly by the indirect production mechanism, as a result, it would be not sensitive to the FNNP scenario.}. As a result, it would be hopeful to exclude or test the FNNP scenario through this Higgs rare decay at the HL-LHC. In this paper, we will consider the next-to-leading order (NLO) QCD correction for the decay width of $H\to \Upsilon(ns)+\gamma$ with the $H\gamma\gamma$ anomalous coupling under the non-relativistic QCD (NRQCD) framework, and  demonstrate below that the FNNP parameter space originated from $H\to \gamma\gamma$ could be excluded or tested by the rare decay $H\to \Upsilon(ns)+\gamma$ at the HL-LHC.

\section{Theoretical analysis} 
In this work we adapt the effective Lagrangian approach to parameterize the possible NP effects and consider the following effective couplings which are related to the Higgs exclusive rare decay,
\beq
\mathcal{L}=\frac{\alpha_{\rm em}}{4\pi v}\left(\kappa_{\gamma\gamma}HA_{\mu\nu}A^{\mu\nu}+\kappa_{Z\gamma}H Z_{\mu\nu}A^{\mu\nu}\right)-\frac{y_b}{\sqrt{2}}\kappa_bH\bar{b}b,
\eeq
where $v=246~{\rm GeV}$ is the vacuum expectation value and $\alpha_{\rm em}$ is the electromagnetic coupling. $Z_{\mu\nu}$ and $A_{\mu\nu}$ are the field strength tensor of the $Z$ boson and photon, respectively. $y_b$ is the bottom quark Yukawa coupling and $\kappa_b$ is introduced to parametrize possible NP effect in the bottom quark Yukawa sector and $\kappa_b=1$ in the SM.
It has been demonstrated that the contribution from $HZ\gamma$ anomalous coupling to the rare decay $H\to\Upsilon(ns)+\gamma$ will be suppressed by $\mathcal{O}(m_b^4/m_Z^4)$~\cite{Konig:2015qat}, therefore,  it is reasonable to ignore its contribution in this work. The $H\gamma\gamma$ anomalous coupling can be well constrained by the $H\to\gamma\gamma$ measurements at the LHC. The partial decay width of $H\to\gamma\gamma$ after including the anomalous coupling $\kappa_{\gamma\gamma}$ is given by,
\beq
\Gamma(H\to\gamma\gamma)=\frac{m_H^3}{16\pi}\left(\frac{\alpha_{\rm em}}{2\pi v}\right)^2 \left(\kappa_{\gamma\gamma}^{\rm SM}+\kappa_{\gamma\gamma}\right)^2,
\eeq
where $\kappa_{\gamma\gamma}^{\rm SM}$ is the contribution from the $W$-boson and top quark loops in the SM~\cite{Djouadi:2005gi},
\beq
\kappa_{\gamma\gamma}^{\rm SM}=\frac{1}{2}\left[3Q_t^2A_t(\tau_t)+A_W(\tau_W)\right].
\eeq
Here the functions $A_t$ and $A_W$ are the contribution from top quark and $W$-boson respectively, and their definitions can be found in Ref.~\cite{Djouadi:2005gi} with $\tau_i=m_H^2/4m_i^2$, $Q_t$ is the electric charge of top quark. The bottom quark loop contribution is suppressed by the Yukawa coupling $y_b$ and is ignored.
The contribution from $W$-boson will dominates over the one from the top quark loop owing to a large beta function coefficient of $W$ loop, which could be understood from the Higgs low energy theorem in the limit of $\tau_i\to 0$~\cite{Low:2009di,Cao:2009ah,Cao:2015scs}, i.e. $A_W(\tau_W)=-8.324$ and $A_t(\tau_t)=1.376$.

The properties of the Higgs boson have been studied extensively in the diphoton final state by the ATLAS and CMS experiments~\cite{ATLAS:2018hxb,ATLAS:2018tdk,ATLAS:2019nkf,CMS:2018uag,CMS:2018ctp,ATLAS:2022tnm}.  The most updated limit is from the ATLAS collaboration at the 13 TeV with $139~{\rm fb}^{-1}$ and the signal strength is $\mu=1.04\pm 0.1$~\cite{ATLAS:2022tnm}. Therefore, we can get the bound for the $H\gamma\gamma$ anomalous coupling at $1\sigma$ level as following,
\beq
0.94\leq \frac{\Gamma(H\to\gamma\gamma)}{\Gamma_{\rm SM}(H\to\gamma\gamma)}\leq 1.14,
\eeq
where the subscript ``SM" indicates the prediction from SM, i.e. $\kappa_{\gamma\gamma}=0$. It yields a bound on $\kappa_{\gamma\gamma}$ as,
\begin{align}
&-0.22\leq \kappa_{\gamma\gamma}\leq 0.10, & 6.39\leq \kappa_{\gamma\gamma}\leq 6.71.
\end{align}
It clearly shows that there are  two-fold solutions for $\kappa_{\gamma\gamma}$ when we only consider the  branching ratio of $H\to \gamma\gamma$ measurements.
Breaking the degenerate solutions of $H\gamma\gamma$ anomalous coupling plays a crucial role to understand the nature of the  Higgs boson. 

Next, we consider the rare decay $H\to\Upsilon(ns)+\gamma$ to probe the FNNP scenario at the HL-LHC.
From the factorization of the NRQCD~\cite{Bodwin:1994jh}, the partial decay width could be written as,
\beq
\Gamma(H\to\Upsilon(ns)+\gamma)=\hat{\Gamma}(H\to (b\bar{b})+\gamma)\langle\mathcal{O}^{\Upsilon(ns)}(^3S_1)\rangle,
\label{eq:fac}
\eeq
where $\hat{\Gamma}(H\to (b\bar{b})+\gamma)$ is the short distance coefficient  and can be calculated from the matching between the perturbative QCD and NRQCD, while the long-distance matrix element $\langle\mathcal{O}^{\Upsilon(ns)}(^3S_1)\rangle$ is a non-perturbative parameter which can be determined either from the lattice QCD  calculation or the branching ratio measurements of $\Upsilon(ns)\to \ell^+\ell^-$, with $\ell^{\pm}=e^{\pm},\mu^{\pm},\tau^{\pm}$. Recently, the complete three-loop QCD corrections to the leptonic decay of $\Upsilon(ns)$ has been finished in Ref.~\cite{Feng:2022vvk}. We also note that the relativistic correction effects will be suppressed by the relative velocity of the bottom quarks in the meson rest frame and its numerical effects is very small and can be ignored as well in this analysis~\cite{Bodwin:2014bpa,Brambilla:2019fmu}. Below, we calculate the partial decay width of $H\to \Upsilon(ns)+\gamma$ under the NRQCD factorization formalism at the leading order (LO) and NLO accuracy of strong coupling $\alpha_s$.

\subsection{LO partial decay width}
As shown in Fig.~\ref{Fig:FeyLO}, there are two different production mechanisms for this exclusive decay $H(p_H)\to \Upsilon(ns)(2p_b)+\gamma(p_\gamma)$. We use the covariant projection operator to calculate the scattering amplitudes, which is defined as,
\beq
\Pi=\frac{\Psi_{\Upsilon(ns)}(0)}{2\sqrt{m_\Upsilon}}\slashed{\epsilon}_\Upsilon^*(p_\Upsilon)\left(\slashed{p}_{\Upsilon}+m_{\Upsilon}\right)\otimes\frac{\bold{1}_c}{\sqrt{N_c}},
\eeq
where $\epsilon_\Upsilon^\mu(p_\Upsilon)$ is the polarization vector of the $\Upsilon$ with the momentum $p_\Upsilon$ and  $\Psi_{\Upsilon(ns)}(0)$ is the Schr\"odinger wave function of the $\Upsilon(ns)$ at the origin which can be related to the long-distance matrix element of Eq.~\eqref{eq:fac} by~\cite{Bodwin:1994jh}
\beq
\Psi_{\Upsilon(ns)}^2(0)=\frac{1}{6N_c}\langle\mathcal{O}^{\Upsilon(ns)}(^3S_1)\rangle.
\eeq
The factor $N_c=3$ and $\bold{1}_c$ in Eq.~\eqref{eq:fac} denotes the unit color matrix. Under the framework of NRQCD, we have $p_\Upsilon=2p_b=2p_{\bar{b}}$ and $m_\Upsilon=2m_b$. In the limit of $m_b\to 0$, the direct-production amplitude of $H\to \Upsilon(ns)+\gamma$ at the LO is,
\begin{align}
M_0^{\rm direct}&=\frac{2\sqrt{3}ey_b\kappa_b\Psi_{\Upsilon(ns)}(0)}{3\sqrt{m_b}m_H^2}\left[m_H^2\epsilon^*_\Upsilon\cdot\epsilon^*_\gamma-4 p_b\cdot\epsilon^*_\gamma p_\gamma\cdot \epsilon^*_\Upsilon\right],
\end{align}
where $e$ is the electron charge and $\epsilon_\gamma^\mu$ is the polarization vector of the photon. The indirect-production amplitude from Fig.~\ref{Fig:FeyLO}(c) is given by,
\begin{align}
M_0^{\rm indirect}&=\frac{\alpha_{\rm em}}{4\pi v}\frac{\sqrt{2}e\Psi_{\Upsilon(ns)}(0)(\kappa_{\gamma\gamma}^{\rm SM}+\kappa_{\gamma\gamma})}{\sqrt{3m_b}m_b}\nn\\
&\times\left[m_H^2\epsilon^*_\Upsilon\cdot\epsilon^*_\gamma-4 p_b\cdot\epsilon^*_\gamma p_\gamma\cdot \epsilon^*_\Upsilon\right].
\end{align}
Combing the amplitudes from the direct and indirect production processes, we obtain the partial decay width of $H\to\Upsilon(ns)+\gamma$ at the LO,
\beq
\Gamma_0=\frac{e^2\Psi^2_{\Upsilon(ns)}(0)}{12\pi m_H m_b}\left[\frac{\alpha_{\rm em}}{4\pi v}\frac{m_H^2}{m_b}\left(\kappa_{\gamma\gamma}^{\rm SM}+\kappa_{\gamma\gamma}\right)+\sqrt{2}y_b\kappa_b\right]^2.
\label{eq:LO}
\eeq
We checked that our result agrees with that in Refs.~\cite{Bodwin:2013gca,Konig:2015qat}. It clearly shows that the interference effect between the direct and indirect production of $H\to \Upsilon(ns)+\gamma$ in the SM ($\kappa_{\gamma\gamma}^{\rm SM}=-3.2445$, $\kappa_{\gamma\gamma}=0$ and $\kappa_b=1$) is destructive, but the partial decay width can be significantly enhanced  within the FNNP parameter space (e.g. $\kappa_{\gamma\gamma}\sim -2\kappa_{\gamma\gamma}^{\rm SM}$).

\subsection{NLO QCD correction}
\begin{figure}
\centering
\includegraphics[scale=0.32]{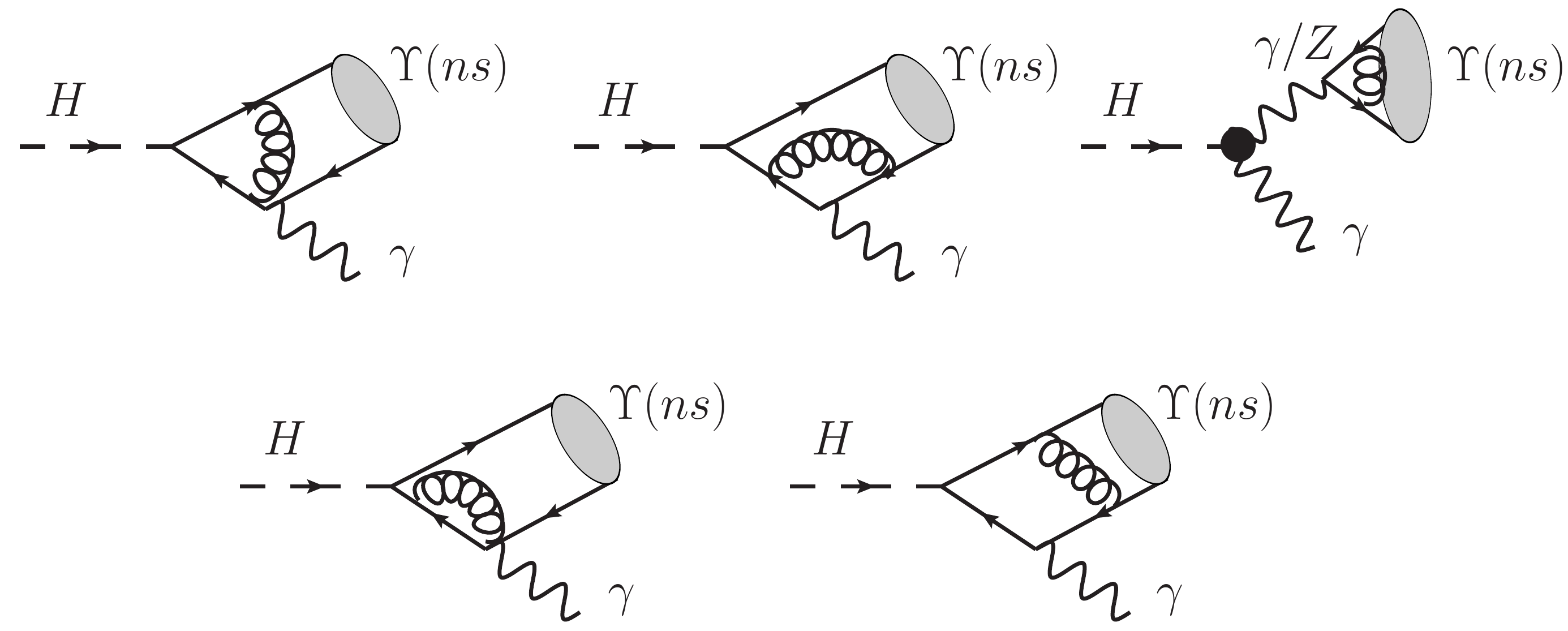}
\caption{Illustrative Feynman diagrams of $H\to \Upsilon(ns)+\gamma$ at the NLO level.}
\label{Fig:FeyNLO}
\end{figure}
Now we consider the NLO QCD correction to the exclusive Higgs decay process $H\to \Upsilon(ns)+\gamma$. The illustrative Feynman diagrams at one-loop level can be found in Fig.~\ref{Fig:FeyNLO}. To regularize the ultraviolet (UV) and infrared (IR) divergences at the loop level, we adopt the dimensional regularization scheme in our calculation and define the dimensionality of space-time  $d=4-2\epsilon$. We should note that the NLO QCD correction contains a Coulomb singularity proportional to $1/v_b$ when the gluon exchanges between the on-shell quark and antiquark. Here $v_b$ is the relative velocity between $b$ and $\bar{b}$ quarks, and it is related to the momenta $p_{b}$ and $p_{\bar{b}}$ by $\overrightarrow{p}_b+\overrightarrow{p}_{\bar{b}}=\overrightarrow{0}$ and $|\overrightarrow{p}_b-\overrightarrow{p}_{\bar{b}}|=m_bv_b$~\cite{Kramer:1995nb}.  After including the renormalization of the loop calculation, we can write the partial decay width of $H\to\Upsilon(ns)+\gamma$ at the NLO level as,
\begin{align}
\Gamma_{\rm NLO}&=\Gamma_0\left(1+\frac{\alpha_s}{\pi}C_F\frac{\pi^2}{v_b}+\frac{\alpha_s}{\pi}F+\mathcal{O}(\alpha_s^2)\right)\nn\\
&\simeq\Gamma_0\left(1+\frac{\alpha_s}{\pi}C_F\frac{\pi^2}{v_b}\right)\left(1+\frac{\alpha_s}{\pi}F\right).
\end{align}
Here the factor $F$ is the finite part of the NLO QCD correction. The Coulomb singularity from box diagrams of direct production and triangle diagram of indirect production has been factorized out and can be absorbed into the definition of the long-distance matrix element $\langle\mathcal{O}^{\Upsilon(ns)}(^3S_1)\rangle$.

To remove the UV divergences, we choose the on-mass-shell (OS) renormalization scheme in this work. The renormalization constants for the quark field and its mass are, respectively,
\begin{align}
\delta Z_2^{\rm OS}&=-C_F\frac{\alpha_s}{4\pi}\left[\frac{1}{\epsilon_{\rm UV}}+\frac{2}{\epsilon_{\rm IR}}-3\gamma_E+3\ln\frac{4\pi\mu^2}{m_b^2}+4\right],\nn\\
\delta Z_m^{\rm OS}&=-3C_F\frac{\alpha_s}{4\pi}\left[\frac{1}{\epsilon_{\rm UV}}-\gamma_E+\ln\frac{4\pi\mu^2}{m_b^2}+\frac{4}{3}\right],
\end{align}
where $\gamma_E$ is the Euler constant, $1/\epsilon_{\rm UV/IR}$ denote the UV/IR poles, and $\mu$ is the renormalization scale. 
After the renormalization procedure,  all the divergences are cancelled. In the limit of $m_b\to 0$, the contributions to the partial decay width originated from the triangle ($\Gamma_{\rm tri}$), self-energy ($\Gamma_{\rm self}$), counter term ($\Gamma_{\rm CT}$) and box diagrams ($\Gamma_{\rm box}$) are, respectively,  
\begin{align}
\Gamma_{\rm tri}&=\frac{\alpha_s}{\pi}C_F\left[3-\frac{\pi^2}{3}-\ln^2 2+\left(\frac{1}{2}+2\ln 2\right)\ln\frac{2m_b^2}{m_H^2}\right.\nn\\
&\left. +\frac{5}{2}\ln\frac{\mu^2}{m_b^2}\right]\left(\Gamma_0^{\rm dir}+\frac{1}{2}\Gamma_0^{\rm int}\right)\nn\\
&+\frac{\alpha_s}{\pi}C_F\left(-2+\frac{3}{2}\ln\frac{\mu^2}{m_b^2}\right)\left(\Gamma_0^{\rm indir}+\frac{1}{2}\Gamma_0^{\rm int}\right),\\
\Gamma_{\rm selft}&=-\frac{\alpha_s}{2\pi}C_F\left(\ln\frac{2\mu^2}{m_H^2}+1\right)\left(\Gamma_0^{\rm dir}+\frac{1}{2}\Gamma_0^{\rm int}\right),\\
\Gamma_{\rm CT}&=-\frac{\alpha_s}{\pi}C_F\left(4+3\ln\frac{\mu^2}{m_b^2}\right)\left(\Gamma_0^{\rm dir}+\frac{1}{2}\Gamma_0^{\rm indir}+\frac{3}{4}\Gamma_0^{\rm int}\right),\\
\Gamma_{\rm box}&=\frac{\alpha_s}{\pi}C_F\left(\ln\frac{4\mu^2}{m_b^2}-2\right)\left(\Gamma_0^{\rm dir}+\frac{1}{2}\Gamma_0^{\rm int}\right),
\end{align}
where $\Gamma_0^{\rm dir}, \Gamma_0^{\rm indir}$ and $\Gamma_0^{\rm int}$ are corresponding to the LO partial decay width from the direct, indirect production and the interference between them, respectively; see the details in Eq.~\eqref{eq:LO}. Note that the Coulomb singularity from box and indirect triangle diagrams has been matched into the long-distance matrix element.  Combing all the parts, we obtain the partial decay width of $H\to \Upsilon(ns)+\gamma$ at the NLO level,
\begin{align}
\Gamma_{\rm NLO}&=\Gamma_0-\frac{\alpha_s(\mu)}{2\pi}C_F\left[7+\frac{2\pi^2}{3}-2\ln^22 \right.\nn\\
&\left. -4\ln 2\left(1+\ln\frac{m_b^2}{m_H^2}\right)\right]\Gamma_0^{\rm dir}\nn\\
&-\frac{\alpha_s(\mu)}{2\pi}C_F\left[\frac{15}{2}+\frac{\pi^2}{3}-\ln^22\right. \nn\\
&\left. -2\ln 2\left(1+\ln\frac{m_b^2}{m_H^2}\right)\right]\Gamma_0^{\rm int}-4\frac{\alpha_s(\mu)}{\pi}C_F\Gamma_0^{\rm indir}.
\end{align}
The NLO QCD correction to the direct production process agrees well with the result in Refs.~\cite{Shifman:1980dk,Wang:2013ywc} in the limit of $m_b\to 0$.
We have numerically checked that the $m_b$ correction to the partial decay width $\Gamma_{\rm NLO}$ is very small and will be ignored in the following numerical analysis.

\section{The $H\gamma\gamma$ anomalous coupling}
Below, we utilize the rare decay of $H\to\Upsilon(ns)+\gamma$ to constrain the $H\gamma\gamma$ anomalous coupling. 
The parameters of the SM are choose as follows~\cite{Workman:2022ynf},
\begin{align}
&m_W=80.385~{\rm GeV}, & m_Z&=91.1876~{\rm GeV}, \nn\\
&m_H=125~{\rm GeV}, & \Gamma_H&=4.07~{\rm MeV}, \nn\\
&G_\mu =1.1663785\times 10^{-5} ~{\rm GeV}^{-2}.
\end{align}
The weak mixing angle is fixed under the $G_\mu$ scheme~\cite{Dittmaier:2001ay}, which is $c_W=m_W/m_Z$ and the electromagnetic coupling $\alpha_{\rm em}=\sqrt{2}G_\mu m_W^2s_W^2/\pi$.
We define the running bottom quark Yukawa coupling $y_b(m_H)$ of the direct production by the running mass at next-to-next-to-leading order in the $\rm \overline{MS}$ scheme, i.e. $m_b(m_H)=2.79~{\rm GeV}$~\cite{Workman:2022ynf}. The long-distance matrix elements $\langle\mathcal{O}^{\Upsilon(ns)}(^3S_1)\rangle$ can be obtained from the branching ratio measurements of $\Upsilon(ns)\to \ell^+\ell^-$ and can be found in Table 1 of Ref.~\cite{Dong:2022ayy}.

\begin{table}
\centering
\caption{The branching ratios of $H\to \Upsilon(ns)+\gamma$ at the LO and NLO, respectively, in unites of $10^{-8}$, with the renormalization scale $\mu=m_H$ and $\kappa_b=1$.} 
\begin{tabular}{|c|c|c|c|}
\hline
${\rm BR}(H\to \Upsilon(ns)+\gamma)$&$\Upsilon(1s)$&$\Upsilon(2s)$&$\Upsilon(3s)$\\
\hline
LO $(\kappa_{\gamma\gamma}=0)$& 0.51 & 0.24 & 0.18\\
\hline
NLO $(\kappa_{\gamma\gamma}=0)$&3.03&1.44&1.05\\
\hline
LO $(\kappa_{\gamma\gamma}=-2\kappa_{\gamma\gamma}^{\rm SM})$& 90.3&42.9&31.1\\
\hline
NLO $(\kappa_{\gamma\gamma}=-2\kappa_{\gamma\gamma}^{\rm SM})$&83.6&39.7&28.8\\
\hline
\end{tabular}
\label{table:BR}
\end{table}

Table~\ref{table:BR} shows the predicted branching ratios of $H\to \Upsilon(ns)+\gamma$ at the LO and NLO with renormalization scale $\mu=m_H$ for the SM and FNNP case ($\kappa_{\gamma\gamma}=-2\kappa_{\gamma\gamma}^{\rm SM}$ and $\kappa_b=1$).  Note that the accuracy of the long-distance matrix elements should be consistent with the  short-distance coefficient when we calculate the branching ratios. It clearly shows that the branching ratios of $H\to\Upsilon(ns)+\gamma$ under the FNNP case could be enhanced about one to two orders of magnitude  compared to the SM predictions.

Searches for the rare decay of $H\to quarkonia+\gamma$ has been performed at the  the ATLAS collaboration at the 13 TeV LHC with an integrated luminosity of $13.6~{\rm fb}^{-1}$~\cite{ATLAS:2018xfc}.
A  95\% confidence level (CL) upper limits on the branching ratios of Higgs decays to $J/\Psi\gamma,\Psi(2s)\gamma,\Upsilon(ns)\gamma$ have been obtained and the upper limits for the $\Upsilon(ns)\gamma$ are $(4.9,5.9,5.7)\times 10^{-4}$. With a much larger integrated luminosity  is collected at the HL-LHC, we could expect that the upper limits of these branching ratios could be improved by two to three orders of magnitude compared to Ref.~\cite{ATLAS:2018xfc}. 
In fact, an estimated projection for the Higgs decays to $J/\Psi\gamma$ has been made by the ATLAS collaboration at the HL-LHC and it shows that the expected 95\% CL upper limit for the branching ratio of $H\to J/\Psi+\gamma$ with $J/\Psi\to\mu^+\mu^-$ is around $\mathcal{O}(10^{-5})$~\cite{ATL-PHYS-PUB-2015-043} . From the analysis of the ATLAS experiment~\cite{ATLAS:2018xfc}, we note that both the signal and background events will be reduced  about  a factor 2  from $H\to \Upsilon(ns)+\gamma$ to $H\to J/\Psi+\gamma$ due to the different quarkonium mass window and cut efficiencies, while the signal efficiency will be enhanced about 20\%.
To estimate the event numbers of the signal $\Upsilon(ns)\gamma$ and background at the HL-LHC, we assume  this relation is still hold and rescale the event numbers from the analysis of Ref.~\cite{ATL-PHYS-PUB-2015-043}, i.e. the event number for $\Upsilon(ns)$ can be obtained by
\beq
n_\Upsilon\simeq 0.6n_{J/\Psi}\frac{{\rm BR}(H\to\Upsilon(ns)+\gamma){\rm BR}(\Upsilon(ns)\to\ell^+\ell^-)}{{\rm BR}(H\to J/\Psi+\gamma){\rm BR}(J/\Psi\to \mu^+\mu^-)}.
\label{eq:event}
\eeq
where $n_{J/\Psi}$ is the event number for the measurement $H\to J/\Psi+\gamma$ at the HL-LHC~\cite{ATL-PHYS-PUB-2015-043}.
Similar to Ref.~\cite{Dong:2022ayy}, we will also combine the measurements from $\Upsilon(ns)$ to $e^+e^-,\mu^+\mu^-$ and $\tau^+\tau^-$ and  assume the same detection efficiency for all three decay channels.

To estimate the sensitivity for testing the hypothesis with parameters $(\kappa_{\gamma\gamma},\kappa_b)$ against the hypothesis with SM, we define the likelihood function as~\cite{Cowan:2010js},
\beq
L(\kappa_{\gamma\gamma},\kappa_b)=\prod_i\frac{(s_i(\kappa_{\gamma\gamma},\kappa_b)+b_i)^{n_i}}{n_i!}e^{-s_i(\kappa_{\gamma\gamma},\kappa_b)-b_i},
\eeq
where $b_i$ and $n_i$ are the event numbers for the background and observed events in the $i$-th process ($H\to\Upsilon(ns)+\gamma$ at the ATLAS and CMS collaborations, with $n=1,2,3$ and $\Upsilon(ns)\to \ell^+\ell^-$), respectively. The parameter $s_i(\kappa_{\gamma\gamma},\kappa_b)$ is the event number of the signal for the parameters $(\kappa_{\gamma\gamma},\kappa_b)$ of the $i$-th data sample, which could be obtained from the rescaling of the $J/\Psi\gamma$ simulation~\cite{ATL-PHYS-PUB-2015-043}; see Eq.~\eqref{eq:event}. Here the label $i$ denotes the experiments from $n=1,2,3,\ell=e,\mu,\tau$, and ATLAS, CMS experiments. The observed events is a combination of the SM signal and background, i.e. $n_i=s_i(\kappa_{\gamma\gamma}=0,\kappa_b=1)+b_i$. The test statistic $q$ is defined as the ratio of the likelihood function,
\beq
q^2=-2\ln\frac{L(\kappa_{\gamma\gamma}\neq 0,\kappa_b\neq 1)}{L(\kappa_{\gamma\gamma}=0,\kappa_b=1)}.
\eeq
Based on the event numbers of signals and backgrounds, we obtain
\beq
q^2=2\left[\sum_i n_i\ln\frac{n_i}{n_i^\prime}+n_i^\prime-n_i\right].
\eeq
Here $n_i^\prime=s_i(\kappa_{\gamma\gamma},\kappa_b)+b_i$ and the 1-$\sigma$ (i.e. 68\% CL) upper limit for the parameter space is setting $q=1$. 

\begin{figure}
\centering
\includegraphics[scale=0.333]{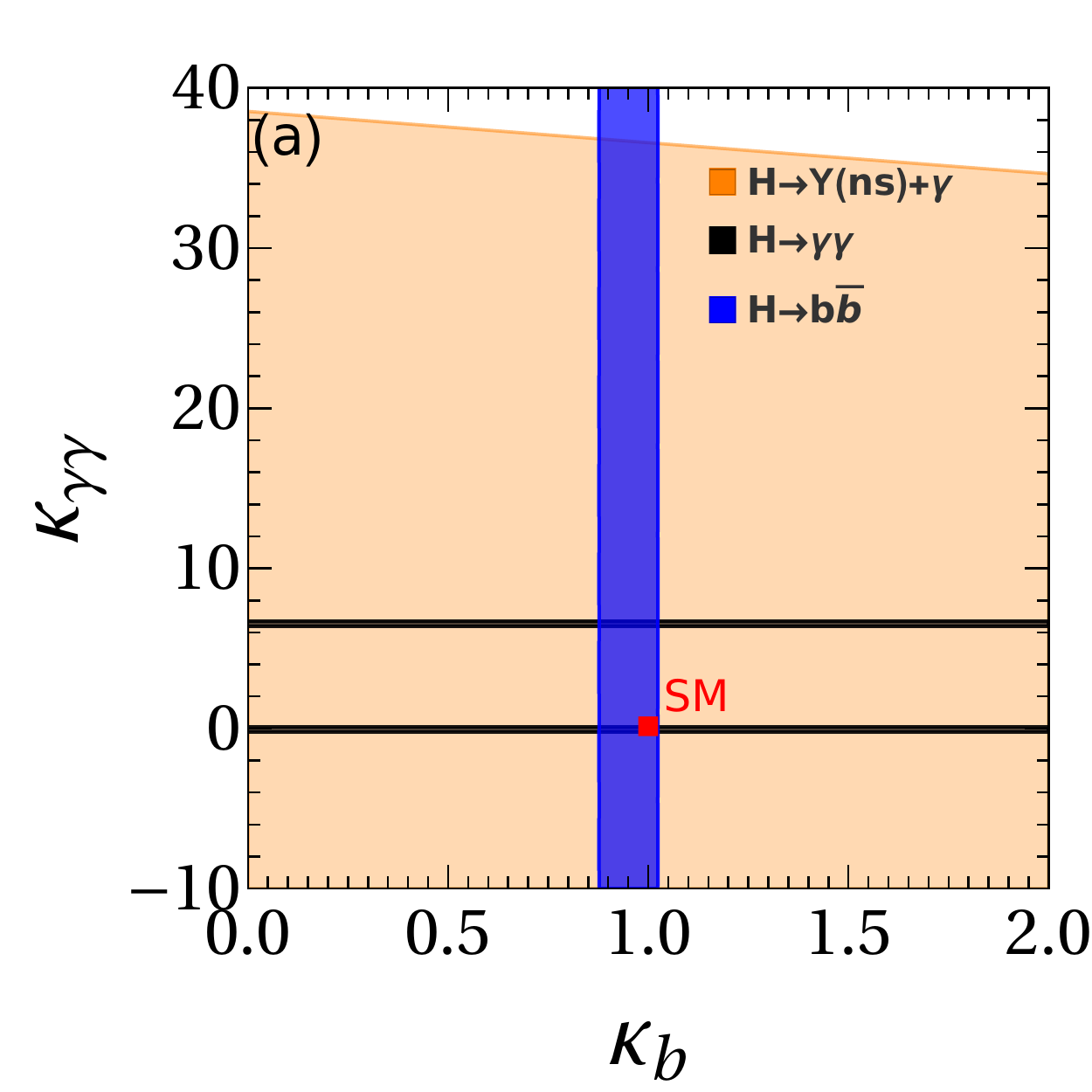}
\includegraphics[scale=0.333]{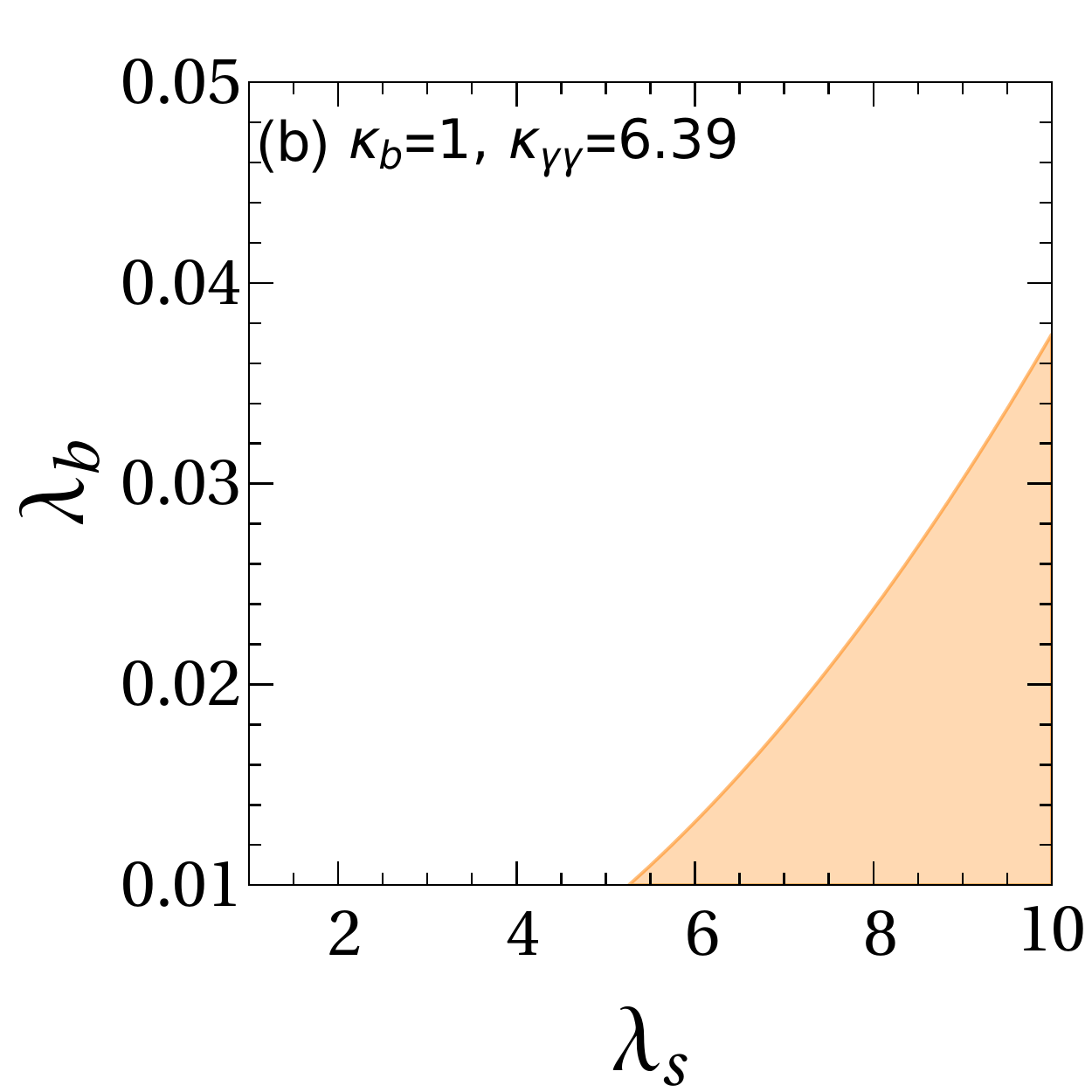}
\caption{(a) The expected 68\% CL limits on the $\kappa_{\gamma\gamma}$ and $\kappa_b$ from the exclusive Higgs decay $H\to\Upsilon(ns)+\gamma\to\ell^+\ell^-+\gamma$ (orange band). The black and blue regions come from the branching ratio measurements of $H\to \gamma\gamma$ and $H\to b\bar{b}$, respectively; (b) The required improvement for the detection efficiencies of background and signal in order to exclude or test the FNNP scenario at the HL-LHC.}
\label{Fig:HL}
\end{figure}

Figure~\ref{Fig:HL}(a) shows the expected 68\% CL constraints on the parameters ($\kappa_{\gamma\gamma},\kappa_b$) obtained from measuring the rare decay $H\to\Upsilon(ns)+\gamma$ at the HL-LHC (orange band). In the same figure, we also show the constrains from the current branching ratio measurements of $H\to\gamma\gamma$~\cite{ATLAS:2022tnm} (black band) and $H\to b\bar{b}$~\cite{ATLAS:2022vkf} (blue band). Though the branching ratios of $H\to\Upsilon(ns)+\gamma$ can be significantly enhanced under the FNNP scenario, it is still very difficult to probe this signal at the HL-LHC based on the event numbers of Ref.~\cite{ATL-PHYS-PUB-2015-043}. 
However, we note that the background analysis from the simulation is evaluated by scaling the observed background at the 8 TeV LHC data and the systematic uncertainty of the background shape is assumed by a constant~\cite{ATL-PHYS-PUB-2015-043}. As a result, the background estimation  from the current analysis should have a large uncertainty and could be overestimated.
With expected advances in the experimental measurement and analysis, it is quite possible that both the signal ($\epsilon_s$) and background ($\epsilon_b$) efficiencies could be improved at the time of HL-LHC runs. We introduce the parameters $\lambda_{b,s}$ to denote the possible improvement of the total efficiencies for the signal and background, respectively, i.e. $\epsilon_{s,b}=\lambda_{s,b}\epsilon_{s,b}^0$, where $\epsilon_{s,b}^0$ denote the efficiencies from the current simulation. We show the required improvement in the detection efficiencies in order to exclude or test  the FNNP scenario in Fig.~\ref{Fig:HL}(b) with $\kappa_b=1$ and $\kappa_{\gamma\gamma}=6.39$. It shows that such goal could be achieved if the background can be suppressed further about two orders of magnitude and the signal efficiency is enhanced about 5 times compared to the ATLAS  preliminary simulation.

\vspace{3mm}
\section {\bf Conclusions}%
In this paper, we propose to probe the $H\gamma\gamma$ anomalous coupling via Higgs exclusive decay $H\to\Upsilon(ns)+\gamma$ at the HL-LHC. 
Owing to the destructive interference between the direct and indirect production mechanisms of the quarkonium, we notice that the branching ratios of $H\to\Upsilon(ns)+\gamma$ could be significantly enhanced under the FNNP  scenario. As a result, it is hopeful to break the degeneracy of the $H\gamma\gamma$ anomalous coupling by this rare decay at the HL-LHC, as implied by the branching ratio of $H\to \gamma\gamma$ measurements at the LHC.
Based on the NRQCD factorization formalism, we calculate the branching ratios of $H\to\Upsilon(ns)+\gamma$ to the NLO accuracy of $\alpha_s$.
To explore the potential of probing the $H\gamma\gamma$ anomalous coupling at HL-LHC, we rescale the background and signal event numbers from the ATLAS simulation for the process $H\to J/\Psi+\gamma$.  It shows that the measurement of Higgs decays to $\Upsilon(ns)\gamma$ could  break the degeneracy of the $H\gamma\gamma$ anomalous coupling, if the background can be further suppressed about two orders of magnitude and the signal efficiency can be improved by a factor of 5, as compared to the current simulation of the ATLAS experiment.

\vspace{3mm}
\noindent{\bf Acknowledgments.}
P. Sun is supported by Natural Science Foundation of China under grant
No. 11975127 and No. 12061131006. BY is supported by the IHEP under Contract No. E25153U1.

\bibliographystyle{apsrev}
\bibliography{reference}

\end{document}